# Evaluating Author Name Disambiguation for Digital Libraries: A Case of DBLP

Jinseok Kim

Institute for Research on Innovation and Science, Survey Research Center, Institute for Social Research, University of Michigan
330 Packard Street, Ann Arbor, MI U.S.A. 48104-2910
jinseokk@umich.edu; jinseok.academic@gmail.com
ORCID ID: 0000-0001-6481-2065

## Abstract

Author name ambiguity in a digital library may affect the findings of research that mines authorship data of the library. This study evaluates author name disambiguation in DBLP, a widely used but insufficiently evaluated digital library for its disambiguation performance. In doing so, this study takes a triangulation approach that author name disambiguation for a digital library can be better evaluated when its performance is assessed on multiple labeled datasets with comparison to baselines. Tested on three types of labeled data containing 5,000 ~ 700K disambiguated names and about 6M pairs of disambiguated names, DBLP is shown to assign author names quite accurately to distinct authors, resulting in pairwise precision, recall, and F1 measures around 0.90 or above overall. DBLP's author name disambiguation performs well even on large ambiguous name blocks but deficiently on distinguishing authors with the same names. When compared to other disambiguation algorithms, DBLP's disambiguation performance is quite competitive, possibly due to its hybrid disambiguation approach combining algorithmic disambiguation and manual error correction. A discussion follows on strengths and weaknesses of labeled datasets used in this study for future efforts to evaluate author name disambiguation on a digital library scale.

# Introduction

Digital libraries, as information systems housing a collection of publication records, are used for various purposes today (Cota, Ferreira, Nascimento, Goncalves, & Laender, 2010). Scholars search digital libraries for conceiving, contextualizing, and designing their research. Commercial digital libraries such as SCOPUS and Web of Science have taken a form of academic intelligence services, providing scholarly performance indicators at individual, institutional, and national levels. Several digital libraries focus on specific science fields such as biomedicine (e.g., PubMed), computer science (e.g., DBLP), and high energy physics (e.g., INSPIRE), while others cover science in general.

A common issue pertinent to most digital libraries is that when users search an author's records by querying her name, it is not always guaranteed that they will get "only and all" results relevant to the target author. Author name ambiguity plays a major role in this information retrieval problem. Specifically, two or more authors happen to share the same name (homonyms), making a digital library to return query outputs not belonging to a target author. In addition, an author may have two or more name variants (synonyms) due to e.g., marriage-induced name change or inconsistent use of middle names, leading to incomplete query outputs. This author name ambiguity in digital libraries may not pose a big problem to casual users who can filter outputs manually for occasional purposes. It can be, however, a crucial issue for those who use digital libraries as a resource for gleaning scientific knowledge to model scholarly activity and gaining practical insights to make informed decision about hiring and funding researchers (Abbott et al., 2010; Hicks, 2012).

Recognizing such negative effects, a few digital libraries such as SCOPUS disambiguate author names algorithmically. Other digital libraries, however, provide authorship information as recorded in their data. Before mining those ambiguous data, therefore, some scholars have resolved name ambiguity using simple heuristics like simplifying name strings in a format of forename initials and surname (e.g., Newman, 2001). Meanwhile, others have applied machine learning algorithms to disambiguate all names in digital libraries (e.g., Torvik & Smalheiser, 2009).

Outcomes of these disambiguation endeavors have been evaluated under different appraisal strategies, sometimes leading to different assessments about the disambiguation performance of the same digital library (e.g., Kawashima & Tomizawa, 2015; Reijnhoudt, Costas, Noyons, Borner, & Scharnhorst, 2014). A proper evaluation of author name disambiguation for digital libraries becomes critical if the validity of research findings mined from the digital library data could be affected by the correctness of author identification in the data. This is also important for consumers of academic intelligence services provided by digital libraries whose products may be compromised by faulty disambiguation.

Considering such importance, this study attempts to evaluate author name disambiguation in DBLP for three reasons. First, as of April 2018, the term "DBLP" appears in almost 700 academic papers indexed in SCOPUS[1]. Many of them have analyzed part of or whole DBLP data for scholarly evaluation, collaboration network modeling, and information retrieval. Findings of those studies may be affected by errors in identifying distinct authors in the DBLP data (Fegley & Torvik, 2013; Kim & Diesner, 2016; Reitz & Hoffmann, 2013). Although DBLP's disambiguation has long been assumed to be accurate by many scholars (Franceschet, 2011), research community other than the DBLP team has not sufficiently tested its disambiguation performance using various labeled data[2]. Second, DBLP team makes publicly available its data monthly. These open data not only enable this study to use them for testing DBLP's disambiguation performance but also allow others to reproduce the results of this study and improve this study's evaluation approach. Third, the evaluation results of DBLP's disambiguation can help scholars make informed decision about their choice of the DBLP data for research in comparison with other digital library datasets.

---

[1] DBLP-related papers were searched using the term "DBLP" with a Document Type filter ("Conference Paper and Article") at https://www.scopus.com/search/form.uri?display=basic

This study evaluates the disambiguation performance of DBLP using multiple labeled data sources. This approach is based on the practice of prior studies for disambiguating author names on a digital library scale. In the following section, related work is introduced to provide context for this study's evaluation method.

## Related Work

A digital library's author name disambiguation (AND hereafter) is, sometimes, considered to be accurate based on the trust of the library's claim (e.g., Ioannidis, Boyack, & Klavans, 2014). Another approach assumes that the true accuracy lies somewhere between upper and lower bounds of disambiguation outcomes generated by two heuristic disambiguation methods (e.g., Newman, 2001). A common practice is, however, to test disambiguation results on labeled data where author name instances are correctly assigned to or grouped for distinct authors by human coders. Several representative studies that used labeled data to assess AND at a digital library level are summarized in Table 1.

In Table 1, two studies evaluated the disambiguation performance of SCOPUS (Kawashima & Tomizawa, 2015; Reijnhoudt et al., 2014). One study (Lerchenmueller & Sorenson, 2016) assessed the disambiguation results by Torvik and Smalheiser (2009) for PubMed data. Meanwhile, other studies disambiguated author names in a whole digital library and measured their performance.

AND studies for digital libraries have relied on different types of labeled data. First, some studies have manually created labeled data (e.g., Wang, Tang, Cheng, & Yu, 2011). For this purpose, scholars collected ambiguous names and publication records associated with the names from a target digital library, and then decided which name belongs to whom by checking information on CVs and personal webpages, or directly asking authors via email (e.g., Levin, Krawczyk, Bethard, & Jurafsky, 2012).

The second type of labeled data has been constructed by matching an author name record in a digital library with author identity information in external data sources. The authority sources are diverse: (1) NIH funding data (Lerchenmueller & Sorenson, 2016; Liu et al., 2014; Torvik & Smalheiser, 2009), (2) national researcher databases in Italy (D'Angelo, Giuffrida, & Abramo, 2011), Japan (Kawashima & Tomizawa, 2015) and The Netherlands (Reijnhoudt et al., 2014), or (3) other digital libraries such as INSPIRE (Louppe, Al-Natsheh, Susik, & Maguire, 2016), SCOPUS (Torvik & Smalheiser, 2009), and the Thompson Reuter's Highly Cited Researcher database (Liu et al., 2014; Torvik & Smalheiser, 2009)[3].

The third type of labeled data has been automatically generated without human curation. Specifically, some scholars assume that if a name in paper X appears in paper Y that cites X, those two name instances appearing both in paper X and Y refer to the same author (Levin et al., 2012; Schulz, Mazloumian, Petersen, Penner, & Helbing, 2014; Torvik & Smalheiser, 2009). Email address has also been used for the automatic construction of labeled data: two names associated with the same email address are assumed to represent the same author (e.g., Schulz et al., 2014).

---

[2] An exception is Kim and Diesner (2015) in which the DBLP's disambiguation performance as of May 2014 was measured on a sample of labeled data (474 distinct authors in 3,921 papers) extracted from Shin, Kim, Choi, and Kim (2014). The evaluation results were *Pairwise F1* = 0.96 and *K-metric* = 0.952.
[3] At the time of this study, the service is provided by Clarivate Analytics at https://clarivate.com/hcr/

As shown in the example studies in Table 1, scholars have evaluated AND for digital libraries using various types of labeled data and evaluation metrics. Two issues are worth noting. Many studies relied on a single labeled dataset. If it is unknown how well a labeled dataset represents the population and name ambiguity in a digital library, reliance on a single labeled dataset is likely to result in incorrect performance evaluation of a digital library's AND. In addition, some studies did not report performance gains compared to baselines, making it difficult to assess properly the effectiveness of the proposed AND method and to compare findings and lessons across studies.

Drawing on the practice of several studies in Table 1, this study takes a multifaceted evaluation approach by testing the AND performance for a digital library on multiple labeled datasets representing various ambiguity levels. Also, this study chooses a widely used heuristic – matching author names on all forename initials – as a baseline performer to capture the performance gains of a digital library's name disambiguation.

*Table 1: Summary of Author Name Disambiguation Studies for Digital Libraries*

| Study | Library (Field) | Record Size | Labeled Data | Performance Measurement |
|---|---|---|---|---|
| Reijnhoudt et al. (2014) | SCOPUS (general) | 45M | 57,775 paper records of 1,400 Dutch professors in National Academic Research and Collaboration Information System (NARCIS) | Precision & Recall<br>Baseline (N/A) |
| Kawashima & Tomizawa (2015) | SCOPUS (general) | N/A | 573,338 paper records linked to 75,405 Japanese researcher IDs in Grants-in-Aid for Scientific Research Database (KAKEN) | Mean Precision & Recall<br>Baseline (N/A) |
| Torvik & Smalheiser (2009) | PubMed (medicine) | 15.3M | (1) Paper pairs of 62 random author names<br>(2) 20,085 author profiles in Community of Science (COS)<br>(3) 2,313 highly cited author profiles in ISI<br>(4) 83,992 PI groups in NIH funding database<br>(5) 323,274 self-citation pairs<br>(6) Random sample of 148 COS author profiles | Lumping & Splitting Errors<br>Baseline (SCOPUS, WOS) |
| Liu et al. (2014) | PubMed (medicine) | 22M | (1) 300 stratified random pairs<br>(2) 40 highly cited researcher profiles<br>(3) 47 NIH researcher profiles<br>(4) 4.7M self-citation pairs<br>(5) 23M grant-citation pairs | (Mean) Pairwise Precision/Recall/F1; Splitting Error<br>Baseline (Torvik and Smalheiser (2009)) |
| Lerchenmueller & Sorenson (2016) | PubMed (medicine) | 15.3M | 355,921 paper records linked to 36,987 principal investigator ids in NIH ExPORTER | Mean Precision & Recall<br>Baseline (N/A) |
| Levin et al. (2012) | Web of Science (general) | 14M | 15,750 paper records of 237 authors who confirmed their publication lists via email | Mean Pairwise Precision/Recall/F1<br>Mean B-Cubed Precision/Recall/F1<br>Baseline (Unsupervised Agglomerative Clustering) |
| Schulz et al. (2014) | Web of Science (general) | 47M | (1) Google Scholar-WOS publication linkage for 3,000 surnames<br>(2) 138 random publication pairs<br>(3) 4.7M name clusters containing at least two names with second forename initials<br>(4) 26,887 clusters of 3,000 random name pairs<br>(5) 110,011 email clusters extracted from arXiv | Mean Precision & Recall<br>Baseline (N/A) |
| Martin et al. (2013) | APS (physics) | 460K | Random selection of 79 matching and 111 nonmatching author pairs with similar names | Merging & Splitting Error<br>Baseline (raw name) |

| Sinatra et al. (2016) | APS (physics) | 450K | Random selection of 200 matching and 200 nonmatching author pairs with similar names | Merging & Splitting Error  Baseline (N/A) |
| --- | --- | --- | --- | --- |
| Wang et al. (2011) | Aminer (computing) | 1.6M | 6,730 paper records of 1,382 authors with same names | Mean Pairwise Precision/Recall/F1  Baseline (DISTINCT, SA-Cluster, CONSTRAINT, HAC) |
| Louppe et al. (2016) | INSPIRE (high-energy physics) | 1M | 360,066 paper records of 15,388 authors identified by original authors, curators, and publishers | Mean Pairwise Precision/Recall/F1  Mean B-Cubed Precision/Recall/F1  Baseline (first-initial method) |

## Target Digital Library

DBLP is a digital library, started by Michael Ley in 1993 to manage metadata of computing research publications and has been widely used by scholars for a variety of research projects (Ley, 2002, 2009). Author names in DBLP are disambiguated in two ways. Note that details on DBLP's disambiguation hereafter are summarized from Müller, Reitz, and Roy (2017). An author name entering DBLP is first compared to names of existing author profiles through several string similarity functions and coauthor information. After manual checking by human experts, one of the profiles is assigned to the new incoming name. In addition, a quality control over the whole DBLP data is performed daily. If two distinct author profiles are found to have a common coauthor (a case of synonym: different in names but referring to the same author), they are merged into one after human inspection. An author profile suspected for a case of homonym (containing publication records of other authors who happen to share the same name) is split into distinct author profiles after scrutiny using a coauthor-community-detection method (e.g., an author profile can be divided into two or more profiles that share no coauthor). Besides these internal algorithmic and manual disambiguation, DBLP allows users to report errors in DBLP records for correction.

For this study, the "September 2017" version of DBLP data was downloaded[4]. A total of 3,746,275 conference or journal paper records were collected for analysis after some publication types such as book, review, and thesis were excluded. DBLP represents distinct authors via name strings. Four-digit numbers are attached to author names to distinguish homonyms; e.g., Bin Liu, Bin Liu 0001, Bin Liu 0002, etc. Sometimes, two or more different names (synonyms) are used for unique authors in DBLP. For these cases, DBLP lists name variants of a distinct author on her DBLP profile under the "aka" (i.e., also known as) section. This study consolidated such synonyms when evaluating DBLP's disambiguation[5].

## Evaluation Methods

For a triangulation of evaluation, this study uses eight labeled datasets obtained from diverse sources. Table 2 summarizes names, sources, types, and sizes of the labeled datasets. Their details are explained in the following sections.

*Table 2: Summary of Labeled Data for DBLP's AND Evaluation*

| Name in this paper | Source | Generation Type | # of Records |
|---|---|---|---|
| PENN | Han et al. (2005) | Manually labeled | 5,018 |
| KISTI | Kang et al. (2011) | | 41,358 |
| AMINER | Tang et al. (2012) | | 6,722 |
| QIAN | Qian et al. (2015) | | 6,567 |
| ORCID | ORCIDs-linked name instances in DBLP | External-authority-linked | 707,137 |
| ORCID Homonym | A set of homonyms from ORCID | | 45,786 |
| ORCID Synonym | A set of synonyms from ORCID | | 62,686 |
| SelfCite | Pairs of name instances in self-citation relation extracted from Aminer | Automatically labeled | 5,969,146 |

---

[4] The dataset in a XML format can be downloaded from dblp.org/xml/release/dblp-2017-09-03.xml.gz
[5] DBLP team kindly provided the list of 39,152 name pairs in synonym relation for this study

**Manually Labeled Data**

To construct labeled data, several studies have obtained seed data (ambiguous names and publication records associated with them) from DBLP and disambiguated author names in the seed data through human inspection. This practice has produced a variety of labeled data derived from DBLP, which can be used in turn to evaluate the performance of DBLP's author name disambiguation. Four hand-labeled datasets originated from DBLP are introduced below. Although other DBLP-induced labeled data (e.g., On, Lee, Kang, & Mitra, 2005; Shin, Kim, Choi, & Kim, 2014) are available, this selection is believed to cover reasonably the most representative hand-labeled datasets for DBLP, which also aligns well with the selection of "most important" labeled data for AND research as reported in Müller et al. (2017).

*PENN*: One of such DBLP-originated labeled data was constructed by C. Lee Giles's team at the Pennsylvania State University in 2004 (Han, Zha, & Giles, 2005) and has been used by many scholars to test disambiguation algorithms (e.g., Cota et al., 2010; Santana, Goncalves, Laender, & Ferreira, 2015). The PENN data were generated by first collecting publication records of ambiguous author names from DBLP and author webpages. Then, researchers determined identities of names by comparing author full name, coauthor name, affiliation, and email address. For this study, a set of 8,453 name record instances (a name record instance in labeled data typically consists of unique author id, ambiguous author name, coauthor name(s), affiliation, title, venue, etc.) was downloaded from Dr. Giles's webpage[6].

*KISTI*: Another labeled data were created by researchers at the Korea Institute of Science and Technology Information in collaboration with the Kyungsung University in Korea (Kang, Kim, Lee, Jung, & You, 2011). The KISTI data are a collection of 41,673 name record instances extracted from 37,613 DBLP-indexed publications. A total of 6,921 unique authors were identified by manual disambiguation exploiting web query results from Google. The original KISTI data were obtained from Santana et al. (2015)[7].

*AMINER*: This labeled dataset was developed by researchers at several Chinese and U.S. institutions under the lead of Jie Tang at the Tsinghua University (Tang, Fong, Wang, & Zhang, 2012; Wang et al., 2011). Ambiguous names in AMINER were gathered from the Arnetminer[8] (a.k.a. Aminer), a computer science digital library indexing publication records collected from DBLP, IEEE, and ACM. Researchers disambiguated author names by comparing information on personal webpages, affiliation, and email address. The downloaded AMINER dataset contains 7,528 name record instances for 110 ambiguous name groups[9].

*QIAN*: The last truth data, QIAN, were labeled by researchers at the Xi'an Jiaotong University in China and the Waseda University in Japan (Qian, Zheng, Sakai, Ye, & Liu, 2015). The data consist of 574 ambiguous name groups in 6,783 name record instances. These data were originally created by combining other labeled data (including PENN and AMINER), and de-duplicated and corrected for errors[10].

Table 3 summarizes the numbers of name record instances and distinct authors in each labeled data after being matched with the downloaded DBLP data. Each record instance in labeled data was compared to DBLP records by comparing year, title (all words were lowercased and non-alphanumeric characters were deleted), venue name, author name (name parts were reordered alphabetically and matched on full or initial strings), and author position, if available. If two or more DBLP records were matching candidates for a record instance, the match was decided by manual inspection via web search of author publication profiles.

---

[6] http://clgiles.ist.psu.edu/data/nameset_author-disamb.tar.zip
[7] http://www.lbd.dcc.ufmg.br/lbd/collections/disambiguation/DBLP.tar.gz/at_download/file
[8] https://aminer.org/
[9] http://arnetminer.org/lab-datasets/disambiguation/rich-author-disambiguation-data.zip
[10] https://github.com/yaya213/DBLP-Name-Disambiguation-Dataset

*Table 3: Numbers of Name Record Instances and Distinct Authors of Four Hand-Labeled Data Before- and After-DBLP-Match*

| Labeled Data | # of Name Record Instances | | | # of Distinct Authors | | |
|---|---|---|---|---|---|---|
| | Before Match | After Match | Match Ratio | Before Match | After Match | Match Ratio |
| PENN | 8,453 | 5,018 | 59.36% | 479 | 480 | 100.21% |
| KISTI | 41,673 | 41,358 | 99.24% | 6,921 | 6,878 | 99.38% |
| AMINER | 7,528 | 6,722 | 89.30% | 1,546 | 1,343 | 86.87% |
| QIAN | 6,783 | 6,567 | 96.82% | 1,201 | 1,182 | 98.42% |

The low match ratio between PENN and DBLP (59.36%) is due in part to erroneous and duplicate record instances in the original PENN data (Müller et al., 2017; Santana et al., 2015; Shin et al., 2014). The error correction during the matching process was conducted at a minimal level but resulted in the increased number of distinct authors (479 → 480) for PENN. Another reason for discrepancy between DBLP and labeled data is that although PENN, AMINER, and QIAN relied on DBLP as their seed data source, they combined other sources (e.g., PENN = online author profiles, AMINER and QIAN = other computing digital library's records) than DBLP. Even KISTI, which imported raw source data solely from DBLP, had mismatch due to its incorrect assignment of author identifiers and DBLP's record updates.

**ORCIDs-Linked Data**

To generate labeled data constructed by an external authority source, this study utilizes ORCIDs, an information system of self-reported authorship profiles[11]. In the ORCIDs system, a registered member is assigned a unique id (ORCID) and its associated publication records are managed by the member herself and, sometimes, added through several metadata providers' contributions (Haak, Fenner, Paglione, Pentz, & Ratner, 2012). In the DBLP data downloaded for this study, a total of 115,843 ORCID ids are connected to 707,137 author name instances. This study uses this set of ORICD id-DBLP author name linkage instances as a *proxy* of truth data for the evaluation of DBLP's AND. Following prior studies, this external-source-based labeled data are assumed to be correct, although the DBLP team is, at the time of writing this paper, validating the accuracy of suspicious cases in DBLP-linked ORCID records[12].

**Automatically Labeled Data**

DBLP author name pairs in self-citation relation (i.e., a name in a paper A appears in a paper B cited by the paper A or vice versa) were detected by a DBLP citation network constructed by the Aminer team[13]. A total of 5,969,146 self-citing name pairs were found by matching author names in citing-cited papers by (1) comparing their first forename initials and surnames and (2) excluding cases where a name in one paper matches two or more names in the other paper, following the common practice of prior studies (e.g., Levin et al., 2012; Liu et al., 2014). As author names in DBLP are recorded in the order of forename plus surname, this study assumes the first and last elements in an author name string separated by space to represent the first forename (first name) and surname (last name), respectively. This simple heuristic can lead to inaccurate surname detection because several ethnic names have two or more surnames. So, the matching results based on this heuristic should be used with this limitation in mind.

---

[11] https://figshare.com/articles/ORCID_Public_Data_File_2017/5479792
[12] For details, see http://dblp.org/faq/17334571
[13] https://static.aminer.org/lab-datasets/citation/dblp.v10.zip

**Baseline Comparison**

Some disambiguation studies have compared their AND methods with baselines or competing algorithms to see how much improvement their proposed method have achieved. In this study, a heuristic – using all available forename initials and full surname to decide author identities – is used as a baseline disambiguation method. According to this method, for example, two names "Mark E. Newman (→ M. E. Newman)" and "Mark Newman (→ M. Newman)" refer to different authors. This initial-based disambiguation method has been widely used in bibliometrics for decades and shown to produce decent to highly accurate disambiguation results (Milojević, 2013). In addition, several digital libraries provide authorship data in this all-initials format. This heuristic disambiguation method can produce bottom-line performance for any AND study for digital libraries because this is applicable across digital libraries at low cost and without complex feature engineering required to run comparative algorithms at a whole library level.

**Measurement**

A suite of pairwise precision/recall/$F$ metrics was consistently used by four studies in Table 1 (Levin et al., 2012; Liu et al., 2014; Louppe et al., 2016; Wang et al., 2011), while other studies used different metrics (Kawashima & Tomizawa, 2015; Reijnhoudt et al., 2014) or proposed their own performance measures (Lerchenmueller & Sorenson, 2016; Martin, Ball, Karrer, & Newman, 2013; Schulz et al., 2014; Sinatra, Wang, Deville, Song, & Barabasi, 2016; Torvik & Smalheiser, 2009). This study uses pairwise-F to make evaluation results comparable across AND studies because this is the most dominantly used metric in disambiguation studies in general (Levin et al., 2012; Menestrina, Whang, & Garcia-Molina, 2010). Pairwise-F measures how accurate AND results are in comparison to labeled data at a pair-level using pairwise-Precision ($p$P), pairwise-Recall ($p$R), and pairwise-F1 ($p$F1) as defined below (Menestrina et al., 2010):

$$pP = \frac{|name\ pairs\ in\ disambiguated\ data\ \cap\ name\ pairs\ in\ labeled\ data|}{|name\ pairs\ in\ disambiguated\ data|} \quad (1)$$

$$pR = \frac{|name\ pairs\ in\ disambiguated\ data\ \cap\ name\ pairs\ in\ labeled\ data|}{|name\ pairs\ in\ labeled\ data|} \quad (2)$$

$$pF1 = \frac{2 \times pP \times pR}{pP + pR} \quad (3)$$

Names without a comparable pair is excluded from calculation. Regarding merging versus splitting, $p$P can be thought of a performance measure for merging errors because two names wrongly paired as referring to the same author (i.e., merged) reduces the value of $p$P by increasing the denominator in equation (1). Likewise, $p$R corresponds to a measure for splitting errors because two names that failed to get paired (i.e., split) leads to low $p$R by decreasing the numerator in equation (2)[14]. The $p$F1 is a harmonic mean of $p$P and $p$R, meaning that merging and splitting are weighed equally. Following the common practice of the referenced studies in Table 1, each pairwise-F measures are calculated per block and then averaged. As a preliminary step to AND, scholars usually group author names into blocks where all names in each block match on the first forename initial and surname. For example, Mark Newman and Mike Newman belong to the same block ('M Newman' Block), but Jake Newman belongs to another block ('J Newman' Block).

---

[14] Splitting can affect the $p$P by decreasing the denominator in equation (1) but also by decreasing the numerator, thus reducing the overall impact of splitting on $p$P.

## Results

### Manually Labeled Data

Table 4 shows the performance of DBLP's disambiguation evaluated on four manually labeled data. Each score is the mean of scores calculated per block and its standard deviation is shown in parentheses. Overall, DBLP's disambiguation is highly accurate considering results tested on PENN, KISTI, and QIAN (close or above 0.90 level). This means, in DBLP data, author names supposed to refer to a unique author were correctly identified in most cases, while some names were wrongly assigned to the author (i.e., merged ≈ lower precision) or others (i.e., split ≈ lower recall).

DBLP is, however, shown to perform poor on AMINER due to a very low precision (= 0.5178). This is because, while other labeled data collected ambiguous names that match on the first forename initial and surname, AMINER collated names that match on the *full* forename and surname but belong to distinct authors (i.e., homonyms). Considering that many AND studies have assumed such homonyms to represent the same author (e.g., Sinatra et al., 2016), the homonym-focused AMINER are believed to be more difficult to disambiguate than other labeled data. DBLP management team has already acknowledged such difficulty of homonym disambiguation (http://dblp.uni-trier.de/faq/1474783) and the DBLP's mediocre precision on the challenging AMINER seems to confirm this.

*Table 4: Mean Performance of DBLP's AND Evaluated on Four Hand-Labeled Data (Mean scores per block reported with standard deviations in parentheses)*

| Labeled Data | Disambiguation Method | Pairwise-F | | |
|---|---|---|---|---|
| | | Mean Precision | Mean Recall | Mean F1 |
| PENN | DBLP | 0.9621 (0.0348) | 0.9576 (0.0433) | 0.9593 (0.0322) |
| | All-Initials | 0.4213 (0.2577) | 0.9844 (0.0138) | 0.5427 (0.2510) |
| KISTI | DBLP | 0.8949 (0.1737) | 0.9693 (0.0873) | 0.9172 (0.1313) |
| | All-Initials | 0.5902 (0.3153) | 0.9797 (0.0688) | 0.6826 (0.2691) |
| AMINER | DBLP | 0.5178 (0.3522) | 0.9328 (0.1338) | 0.5924 (0.3192) |
| | All-Initials | 0.4144 (0.3221) | 0.9865 (0.0635) | 0.5140 (0.3141) |
| QIAN | DBLP | 0.9464 (0.1927) | 0.9634 (0.1764) | 0.9507 (0.1837) |
| | All-Initials | 0.7387 (0.3376) | 0.9702 (0.1701) | 0.7949 (0.2862) |

According to Table 4, DBLP performed better on recall (≈ less splitting) than precision (≈ less merging). For PENN and QIAN, however, the performance differences between precision and recall (|precision − recall|) are within a margin of 0.02, indicating that DBLP produced quite balanced disambiguation results in terms of precision and recall. In contrast, the precision-recall differences are wider for KISTI (0.07) and, most starkly, for AMINER (0.42), meaning that DBLP performed worse on precision than recall. These observations imply that depending on the choices of labeled data, the performance of an AND method can be evaluated in different ways.

Another ambiguity dimension other than merging versus splitting is block size, i.e., the number of name instances to be disambiguated together. As name ambiguity is assumed to increase with block size (Levin et al., 2012; Torvik & Smalheiser, 2009), a good disambiguation method would perform consistently even on large blocks. Figure 1 reports the mean F1 of DBLP's disambiguation (left $y$-axis) per block size for each labeled data with cumulative ratios of block sizes (right $y$-axis).

Except for PENN, a majority of blocks in each labeled dataset are small, while a few blocks are very large. For example, blocks containing 1, 2, or 3 names constitute more than 50% of all blocks in QIAN. Regardless of block sizes, however, DBLP's disambiguation tested on PENN, KISTI, and QIAN shows high performance hovering around F1 of 0.90. Visually, F1 data points are densely clustered towards the ceiling (1.00 level). In contrast, on AMINER, DBLP's performance per block size varies widely: the data points are scattered both vertically and horizontally, meaning that DBLP disambiguated well names in some blocks but not in others and such inconsistent performance happened across most block sizes in AMINER.

The performance gains by DBLP's disambiguation are more clearly observed when compared to baseline performance by the all-initials-based disambiguation. Back in Table 4, DBLP improved overall performance substantially from baselines through the increased precision (+0.1034 on AMINER ~ +0.5408 on PENN) at the cost of small recall losses (−0.0068 on QIAN ~ −0.0537 on AMINER). These gains by DBLP are confirmed in Figure 1. Data points in baseline subfigures (Figure 1: e, f, g, h) spread widely across both $x$-axis and $y$-axis. On PENN, KISTI, and QIAN, especially, they are densely gathered around the upper-left corners, spreading out towards the lower-right corners, meaning that the baseline performance deteriorated as the block size increased. In contrast, subfigures for DBLP (Figure 1: a, b, d) show that data points are densely clustered upwards across block sizes, implying that DBLP disambiguated successfully many ambiguous names, even in large blocks, which the baseline performer failed to disambiguate correctly. For AMINER, DBLP perform better than the all-initials method (i.e., some DBLP's data points are positioned higher than those of All-Initials) but its performance gains are not much pronounced as those on other labeled data.

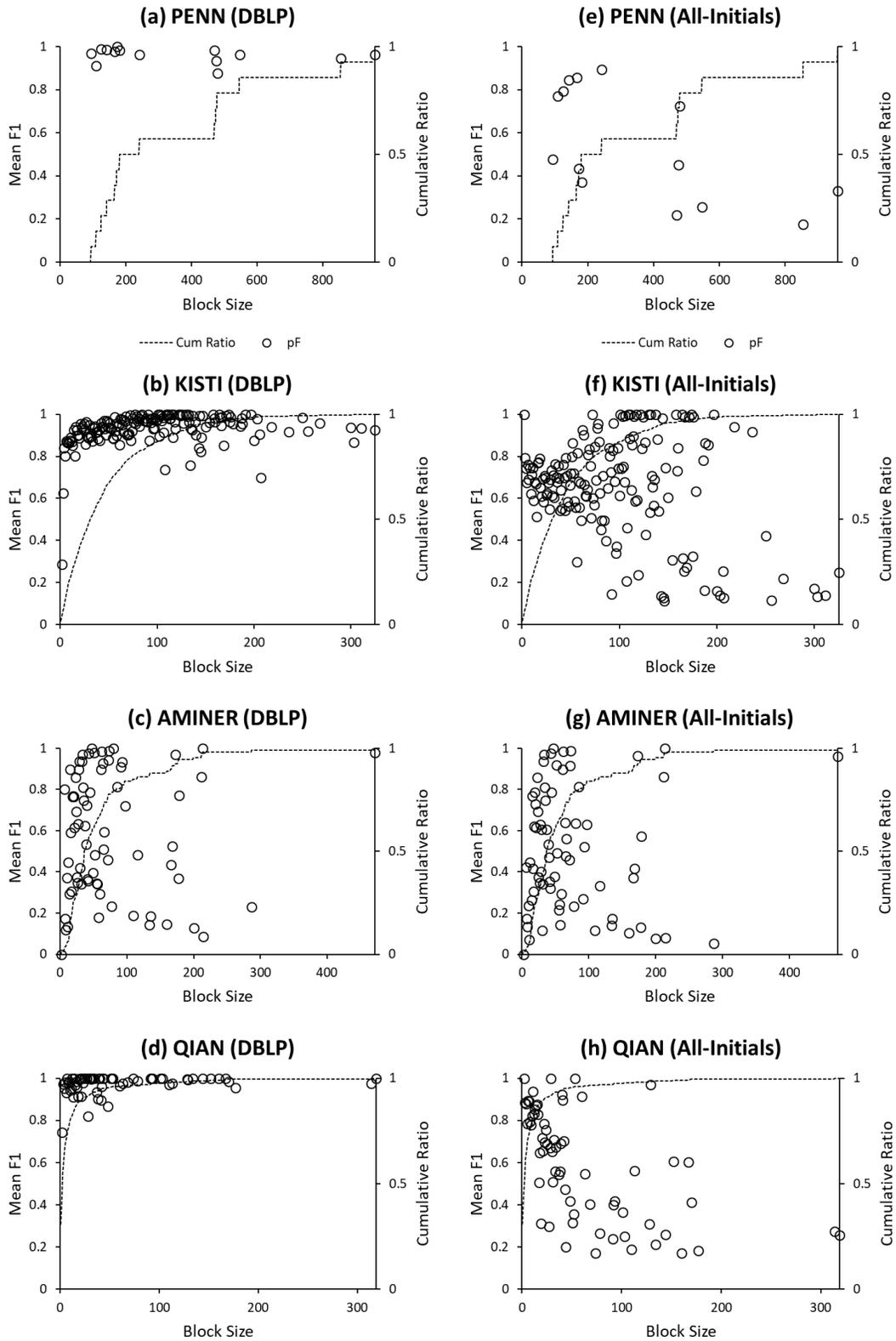

*Figure 1: Mean F1 of DBLP's AND Per Block Size on Hand-Labeled Data (Mean F1 on the left y-axis, Cumulative ratio of block sizes on the right y-axis)*

**ORCIDs-Linked Data**

More than 700K author names in DBLP are associated with ORCID ids of 115,843 distinct authors. This study uses this large linked dataset as a proxy of labeled data (ORCID hereafter). Table 5 reports the evaluation results of DBLP's AND on these data. Overall, DBLP's performance looks almost perfect: precision, recall, and F1 scores are above or close to 0.99. In contrast to the results tested on four hand-labeled data, DBLP performed better on precision than on recall, although the performance differences between precision and recall are within 0.01 level. The high performance of DBLP, however, becomes less outstanding when compared to the baseline score, which is slightly lower than DBLP's. This implies that many DBLP author names associated with ORCID ids are identifiable by matching name strings on forename initials and surname.

To understand deeper the DBLP's performance, name record instances of ORCID were sub-sampled to generate two types of ambiguous name sets. The first was generated on the criteria that (1) names are linked to ORCID ids, (2) they match with at least one other name, and (3) they belong to blocks representing two or more distinct authors. The outcome was a collection of 45,786 homonymous names associated with 8,488 distinct authors (ORCID Homonym hereafter). On this homonym set, DBLP's performance was still high in recall (mean $pR = 0.9828$) but produced merging errors (mean $pP = 0.8512$). The precision gains compared to the baseline are, however, not negligible (+0.1122), while recall gains are very small (+0.0020).

Another subset consists of DBLP names that (1) are linked to ORCID ids, (2) do not match with at least one other name, and (3) belong to blocks representing single distinct authors. This selection resulted in synonym-focused labeled data (ORCID Synonym hereafter) where 3,802 distinct authors are represented by two or more name variants among 62,686 name instances. Tested on this dataset, DBLP's disambiguation recorded near-perfect precision (0.9997) but failed to attach some name instances to distinct author clusters where they should belong (0.9207). Notably, all-initials method produced higher recall than DBLP for synonyms, implying that despite spelling variants, the selected synonym cases match on all forename initials and full surname in most cases.

*Table 5 : Mean Performance of DBLP's AND Evaluated on ORCIDs-linked Labeled Data (Mean scores per block reported with standard deviations in parentheses)*

| Labeled Data | Disambiguation Method | Pairwise-F | | |
|---|---|---|---|---|
| | | Mean Precision | Mean Recall | Mean F1 |
| ORCID | DBLP | 0.9965 (0.0391) | 0.9888 (0.0716) | 0.9903 (0.0566) |
| ORCID | All-Initials | 0.9594 (0.1489) | 0.9857 (0.0796) | 0.9617 (0.1260) |
| ORCID Homonym | DBLP | 0.8512 (0.2344) | 0.9828 (0.0773) | 0.8888 (0.1752) |
| ORCID Homonym | All-Initials | 0.7390 (0.3016) | 0.9808 (0.0816) | 0.7980 (0.2475) |
| ORCID Synonym | DBLP | 0.9997 (0.0117) | 0.9207 (0.1804) | 0.9465 (0.1287) |
| ORCID Synonym | All-Initials | 0.9916 (0.0612) | 0.9522 (0.1427) | 0.9638 (0.1069) |

Like the case of hand-labeled data, DBLP's performance per block size was tested on ORCID. According to Figure 2-a, block size distribution is heavily skewed: almost 83% of all blocks contained 10 or less names, while block sizes ranged from 1 to 838. The F1 data points densely clustered towards the ceiling across most x-axis values mean that DBLP disambiguated quite well regardless of block sizes. This consistent performance is not seen for the all-initials method: data points are densely clustered in the upper-left corner and spread towards the lower-right corner (Figure 2-d). The skewed distribution of block sizes and F1 score spread, when combined, can explain the *small* performance gaps between DBLP's disambiguation and all-initials method shown in Table 5. As small blocks constitute the majority of all blocks in ORCID, the good performance of all-initials method on small-size blocks raised its overall performance (i.e., F1 scores averaged over all block sizes) to as close as the DBLP's performance level. According to the zoomed-in evaluation per block size in Figure 2-a-d), however, the performance of all-initials method could not par DBLP's on large blocks.

Performance test on two ORCID subsets shows similar patterns as that on ORCID. Note that precision is reported for test results on ORCID Homonym because we are interested in how well DBLP's AND distinguishes name instances that share the same name strings but belong to different authors. Likewise, recall is reported for ORCID Synonym because we are interested in how well DBLP's disambiguation method clusters different names that belong to a distinct author. On most block sizes, DBLP showed decent (0.7 ~ 0.9) or high (0.9 or above) levels of precision (Figure 2-b) and recall (Figure 2-c). In contrast, the all-initials method performed worse than DBLP's on most homonym cases (Figure 2-e). On average, however, the baseline performer produced better disambiguation results on synonyms (Table 5 ORCID Synonym) but with wider variations in recall scores than DBLP's across block sizes (Figure 2-f).

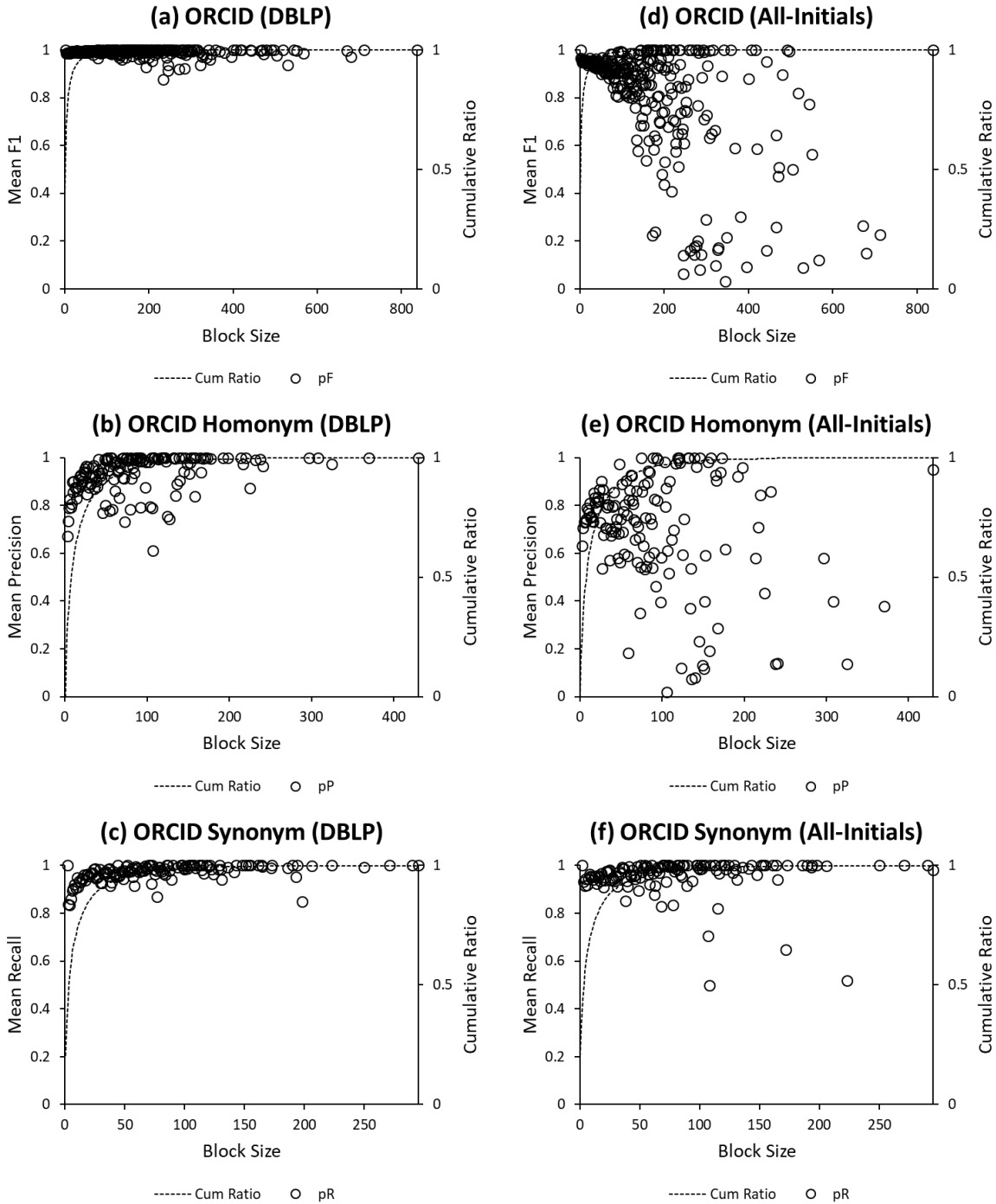

*Figure 2: Mean Performance of DBLP's AND Per Block Size on ORCIDs-linked Labeled Data (Mean F1, Precision, and Recall on the left y-axis, Cumulative ratio of block sizes on the right y-axis)*

**Automatically Labeled Data**

The performance of DBLP's disambiguation was tested on automatically labeled data (SelfCite hereafter) consisting of pairs of name instances in self-citation relation. These labeled data can be used for measuring recall defined as the ratio of the number of pairs matched by DBLP's disambiguation over the number of pairs in the labeled data (Liu et al., 2014; Schulz et al., 2014; Torvik & Smalheiser, 2009). As this recall metric considers only name pairs in self-citation relation, it is different from the pairwise-F recall which considers all possible pairs among name instances referring to the same author.

Out of nearly 6M pairs in the labeled data, DBLP correctly disambiguated on average 98.94% (SD = 0.0776) of them. This score looks very high but less impressive when compared to the recall by the all-initials method (mean recall = 0.9914, SD = 0.0731). When recall was tested per block size (see Figure 3), DBLP's performance showed some variations in recall until around the block size of 2,000 but stabilized on larger blocks.

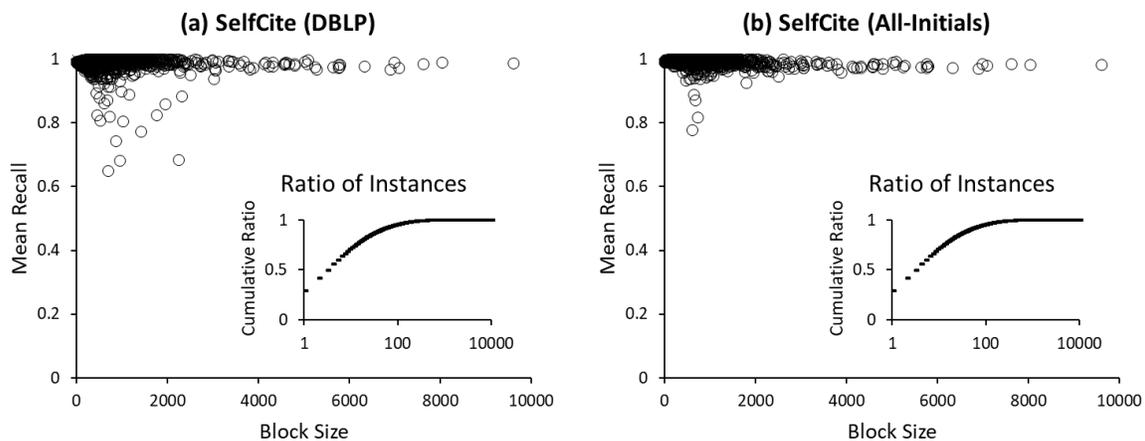

*Figure 3: Mean Recall of DBLP's AND per Block Size on Self-Citation Labeled Data (Inset Figure for Ratio of Instances per block size)*

## Conclusion and Discussion

This study showed how author name disambiguation (AND) for a digital library can be evaluated from a triangulation approach with the case of DBLP. The DBLP's disambiguation performance could be better understood by using various labeled data, considering ambiguity dimensions such as block size, and being compared to baselines. Tested on various labeled data, DBLP's name disambiguation scored higher on recall (≈ less splitting) than on precision (≈ less merging) in most cases (see Table 4 and Table 5). Compared to baselines, however, DBLP's performance gains mostly came from increased precision than improved recall, supporting the claim that DBLP has made more efforts towards high precision than high recall (Müller et al., 2017).

Regarding DBLP's high recall, the evaluation results are limited in telling how well DBLP disambiguated synonymously ambiguous (≈ splitting-error-prone) names. The main reason is that hand-labeled data in this study were generated by screening author names that match on first name initial + surname or full forename + surname, leading to over-sampling of homonym cases (Müller et al., 2017) which is presumed to produce high recall. In addition, the all-initials-based disambiguation in this study was conducted on the DBLP data where name strings have already been disambiguated and expanded (e.g., if two different names – "Mark Newman" and "Mark E. Newman" – are decided to refer to the same author, "Mark Newman" is converted into "Mark E. Newman") (Müller et al., 2017). This implies that part of high recall as well as precision by all-initials method might be due in part to the DBLP's disambiguation.

Across multiple labeled datasets, DBLP's disambiguation showed consistent performance on large blocks, except for AMINER. Notable is that the all-initials method also disambiguated well some large blocks (some data points for All-Initials are positioned close to 1.00 in figures), although it failed to do so for some small blocks. This indicates that some large blocks, e.g., containing names of a prolific author whose names are unique, can be less ambiguous than small blocks that contain name instances of several distinct authors with common names. This also implies that depending on the characteristics of name ambiguity (e.g., author names other than highly ambiguous Asian names are dominant in data), the simple initial-based disambiguation can produce highly accurate results even for large bibliographic data (Milojević, 2013).

Other than the all-initials method, DBLP's AND performance can be compared to other advanced algorithms that have been tested on the same four hand-labeled data used in this study. Results[15] in Table 6 showed that DBLP's disambiguation is quite competitive when compared to top-notch supervised and unsupervised machine learning techniques, although it underperformed on AMINER. Note that these performances were evaluated under different conditions. First, DBLP's performance was tested on the same but slightly smaller labeled data (see Table 3). Second, disambiguation algorithms tested on PENN, KISTI and QIAN were implemented using basic metadata, while DBLP's disambiguation has been enhanced by manual inspection by DBLP team and error correction by authors. In contrast, the AMINER study, unlike other studies as well as DBLP, trained algorithms on enriched feature information such as affiliation, online profile, and citation information. Even the same algorithm (e.g., DISTINCT in AMINER and QIAN) produced different performances depending on labeled data for training and hyper-parameter settings.

*Table 6: Performance Comparison of Various Disambiguation Approaches on Hand-labeled Data (SD in parentheses, if available; performance scores for AMINER were estimated by measuring bar heights in Figure 2 in Wang et al. (2011) using the GNU Image Manipulation Program)*

| Labeled Data | Reference | Measure | Disambiguation Method | Precision | Recall | F1 |
|---|---|---|---|---|---|---|
| PENN | Table 2 in Santana et al. (2015) | Pairwise-F | Naïve Bayes | – | – | 0.616 (0.080) |
| | | | SVM | – | – | 0.702 (0.070) |
| | | | SLAND | – | – | 0.869 (0.034) |
| | | | Nearest Cluster | – | – | 0.919 (0.023) |
| | | | **DBLP** | – | – | **0.959 (0.032)** |
| KISTI | Table 3 in Santana et al. (2015) | Pairwise-F | Naïve Bayes | – | – | 0.596 (0.009) |
| | | | SVM | – | – | 0.623 (0.009) |
| | | | SLAND | – | – | 0.806 (0.008) |
| | | | Nearest Cluster | – | – | 0.816 (0.009) |
| | | | **DBLP** | – | – | **0.917 (0.131)** |
| AMINER | Figure 2 in Wang et al. (2011) | Pairwise-F | CONSTRAINT | 0.87 | 0.60 | 0.70 |
| | | | DISTINCT | 0.90 | 0.60 | 0.72 |
| | | | SA-Cluster | 0.81 | 0.71 | 0.76 |
| | | | HAC | 0.92 | 0.72 | 0.81 |
| | | | ADANA | 0.96 | 0.85 | 0.89 |
| | | | **DBLP** | **0.52** | **0.93** | **0.59** |

---

[15] For comparison, decimal points of performance results in Table 4 were modified to be consistent with metric units in other studies. Also, B-Cubed metrics were calculated for QIAN on all names regardless of block size, following the referenced study.

| | | | K-Means | 0.7812 | 0.5738 | 0.6616 |
| QIAN | Table 12 in Qian et al. (2015) | B-Cubed | DBSCAN | 0.8754 | 0.8111 | 0.8420 |
| | | | DISTINCT | 0.8492 | 0.8513 | 0.8596 |
| | | | BatchAD | 0.8730 | 0.8637 | 0.8683 |
| | | | **DBLP** | **0.9758** | **0.9925** | **0.9841** |

In conclusion, the evaluation results reported in this paper suggest that scholars can regard DBLP data as highly accurate in disambiguating author names. But a caveat to keep in mind is that some homonym cases (distinct authors with the same names) may not be properly distinguished. As DBLP team is testing new algorithmic approaches such as a multilayer perceptron for homonym disambiguation, the accuracy of DBLP's disambiguation on homonyms is expected to improve over time (Ackermann & Reitz, 2018; Momeni & Mayr, 2016). Besides the empirical evaluation of DBLP's AND, another takeaway of this study is that a more granulated, comprehensive understanding of disambiguation performance for digital libraries can be obtained through the performance test triangulated on multiple labeled data and baseline comparison.

Thanks to previous studies manually labeling authorship data derived from DBLP, this study could use various hand-labeled data for evaluating DBLP's AND. However, they are not always attainable for digital libraries other than DBLP. Manual labeling is labor-intensive and time-consuming. In addition, although most hand-labeled data aim to collate most ambiguous name cases, merging-prone cases tend be more prevalent in them than splitting-susceptible ones, which can result in incomplete evaluation of a name disambiguation method (Müller et al., 2017).

As an alternative to the costly hand-labeled data, this study made a proxy of labeled data by collecting author names in DBLP that are linked to ORCID ids. This external-authority-linking can result in large-scale labeled data, which hand-labeling can seldom produce. In addition, this study demonstrated that from the ORCIDs-linked labeled data, we can create stratified samples of names that are susceptible to specific disambiguation errors, merging or splitting, and thereby can help us better measure the DBLP's disambiguation performance. It is, however, unclear how well author records in ORCIDs can represent the population of authors in DBLP (Lerchenmueller and Sorenson, 2016). Moreover, the accuracy of ORCIDs records has not yet been properly evaluated.

The third type of labeled data was constructed by several millions of name pairs in self-citation relation. The scale has a potential to produce a representative sample to test a digital library's AND. This automatic labeling approach can, however, only measure disambiguation performance in a partial way. The matching pairs cannot tell how well a disambiguation method match (≈ recall) or distinguish (≈ precision) pairs of names if they appear in one of matching pairs in labeled data but are not directly in matching relation. Most importantly, accuracy of matching pairs, i.e., whether they really refer to the same authors or not, is not always certain.

As such, this study illustrates how the disambiguation performance of a digital library can be evaluated and provides lessons for future efforts to replicate this study's approach to other digital libraries. A better evaluation of a digital library will be possible after several issues for labeled data such as the coverage of external authority data, unbiased subsampling, and accuracy of automatic matching methods are properly addressed.

Acknowledgements

I would like to thank Florian Reitz (Leibniz Center for Informatics, Schloss Dagstuhl, Germany) for providing the list of synonyms in DBLP and Alan Filipe Santana (Departamento de Ciência da Computação, Universidade Federal de Minas Gerais, Brazil) for sharing the raw KISTI dataset. I am also thankful to anonymous reviewers for their comments. This work was supported by grants from the National Science Foundation (grants #1561687 and #1535370), the Alfred P. Sloan Foundation and the Ewing Marion Kauffman Foundation.